\documentclass[prb,reprint,superscriptaddress,floatfix,twocolumn,showpacs]{revtex4-1}
\pdfoutput=1
\usepackage{graphicx}
\usepackage{amsmath}

\usepackage{color}
\usepackage{soul}

\begin{document}

\title{Unconventional Features in the Quantum Hall Regime of Disordered
Graphene: Percolating Impurity States and Hall Conductance Quantization}

\author{Nicolas Leconte}
\affiliation{Catalan Institute of Nanoscience and Nanotechnology (ICN2), CSIC and The Barcelona Institute of Science and Technology, Campus UAB, Bellaterra, 08193 Barcelona, Spain}
\affiliation{Department of Physics, University of Seoul, Seoul 130-742, Korea}
\email{lecontenicolas@gmail.com}
\author{Frank Ortmann}
\affiliation{Institute for Materials Science and Dresden Center for Computational Materials Science, Technische Universit\"at Dresden, 01062 Dresden, Germany}
\author{Alessandro Cresti}
\affiliation{Univ.  Grenoble Alpes, IMEP-LAHC, F-38000 Grenoble, France}
\affiliation{CNRS, IMEP-LAHC, F-38000 Grenoble, France}
\author{Stephan Roche}
\affiliation{Catalan Institute of Nanoscience and Nanotechnology (ICN2), CSIC and The Barcelona Institute of Science and Technology, Campus UAB, Bellaterra, 08193 Barcelona, Spain}
\affiliation{ICREA - Institucio Catalana de Recerca i Estudis Avan\c cats, 08010 Barcelona, Spain}
\date{\today}

\bibliographystyle{apsrev4-1}

\begin{abstract}
We report on the formation of critical states in disordered graphene, at
the origin of variable and unconventional transport properties in the
quantum Hall regime, such as a zero-energy Hall conductance plateau in
the absence of an energy bandgap and Landau level degeneracy breaking.
By using efficient real-space transport methodologies, we compute both
the dissipative and Hall conductivities of large size graphene sheets
with random distribution of model single and double vacancies. By
analyzing the scaling of transport coefficients with defect density,
system size and magnetic length, we elucidate the origin of anomalous
quantum Hall features as magnetic-field dependent impurity states, which
percolate at some critical energies. These findings shed light on
unidentified states and quantum transport anomalies reported
experimentally.

\end{abstract}

\maketitle
\section{Introduction} 

The role of disorder in the Quantum Hall Effect (QHE)~\cite{Klitzing}  has been essentially related to the existence of a localization/delocalization transition between electronic states, with the formation of critical (extended) states at the center of Landau Levels (LLs)~\cite{Aoki:1987do}. 
In very clean samples, the presence of this transition is ensured by the system edges, which force the formation of extended states, while bulk states are localized by the magnetic field. 
The robustness of the QHE in the bulk limit is guaranteed by the contribution of either weak impurity potentials satisfying the so-called \textit{weakness condition}~\cite{Thouless1981,Halperin1982}, strong scattering centers sufficiently far away from each other~\cite{Prange1982,Chalker1984,Joynt1984}, or smooth potentials with long-range spatial variation~\cite{Joynt1984,Iordansky1982,Floser2013,Floser2013,Prange1982b,Kazarinov1982,Luryi1983,Trugman1983,Giuliani1983,Tsukada1976}. Whenever disorder becomes too strong, all QHE features eventually vanish away. 

Under high enough magnetic fields, the electronic properties of graphene are characterized by the presence of a four-fold degenerate zero energy LL (where electrons and holes coexist) together with non-equidistant LLs at energies $E_{n}=sgn(n)\sqrt{2\hbar{v_{F}}^{2}eB|n|}$ ($v_{F}$ is the Fermi velocity, $B$ is the magnetic field and $n$ is the integer LL index)~\cite{McClure:1956vp,CastroNeto:2009cl,Novoselov2005197,Zhang2005201,Goerbig:2011bh}.  This electronic spectrum results in a Hall conductance quantization $\sigma_{\text{xy}}=\frac{4e^{2}}{h}(n+\frac{1}{2})$~\cite{Goerbig:2011bh}, which is weakly affected by electron-hole puddles or weak surface disorder, but can exhibit further fragmentation of the plateau structure whenever additional symmetry-breaking mechanisms lift the four-fold degeneracy~\cite{Rickhaus:2012br,Barlas:2012et}. 
The presence of an additional quantized Hall plateau $\sigma_{xy}=0$ at low energy in high-mobility samples has been for instance assigned to Zeeman splitting or to the formation of quantum Hall ferromagnetism~\cite{Zhang:2006hn,Nomura:2006bi,Young:2012bn}, with B-field dependent transport scaling behavior conveyed by the dominant symmetry breaking mechanism at play~\cite{Jung:2009hm,Young:2012bn}.

Recently, several experiments have reported unexplained QHE features in disordered graphene, including sets of extended states in Hall measurements and the formation of a zero energy Hall plateau~\cite{PhysRevB.90.195433,Li:2007ju,Nam:2013ey,Nam2014,Jung:2011dk}. These features do not fit the usual energy quantization scheme of massless and massive Dirac charge carriers, and are, as such, often generically attributed to disorder. 
Additionally, the observation of a quantized Hall conductance in highly resistive (millimeter-scale) hydrogenated graphene, with mobility less than $10 {\rm cm}^{2}/{\rm V.s}$ and estimated mean free path far beyond the Ioffe-Regel limit~\cite{Ioffe1960237}, suggests some unprecedented robustness of the QHE in damaged graphene~\cite{GUI_PRL110}.

These findings are considered unconventional in the sense that common belief often states that high concentrations of strong disorder are detrimental to the Hall quantization in 2DEGs. In graphene, however, a variety of literature suggests that things are different and more subtle. The topological contribution to the Berry phase~\cite{Xiao_2010}, which is basically a \textit{winding number} of the pseudo-spin $1/2$~\cite{Fuchs_2010, Park_2011}, is predicted to be more robust under disorder, as it should persist even in the presence of sub-lattice symmetry breaking and associated gap-opening. Such behavior has been demonstrated experimentally in hydrogenated graphene~\cite{PhysRevB.92.125410}. Also, a robust QHE is expected in graphene, even when strong impurities are at a distance smaller than the magnetic length from each other. For instance, for dense impurity concentrations, depending on the symmetry class and impurity strength, a splitting of the critical energy within a single Landau level is predicted when disorder introduces valley mixing~\cite{PhysRevB.75.033412,PhysRevB.76.205408,OST_PRB77,PhysRevLett.112.026802, PhysRevLett.101.036805}, similar to splittings already discussed for 2DEG under the influence of certain types of smooth potential~\cite{PhysRevB.50.7743}. In (quasi-)periodic systems, impurity-engineered Landau levels have been proposed to exist as well~\cite{PER_PRB78}.

Up to now, a proper quantitative description of these phenomena for a completely random distribution of disorder and different types of disorder has been lacking, mostly due to computational limitations. 


In this Article, by using efficient computational methods, we provide tangible numerical insight into the rich physics of the QHE in disordered graphene, backing up the single-particle scenarios~\cite{PhysRevB.75.033412,PhysRevB.76.205408,OST_PRB77,PhysRevLett.112.026802,PER_PRB78, PhysRevLett.101.036805} in which dense distributions of defects can explain the formation of a zero energy plateau in disordered graphene~\cite{Nam:2013ey,Nam2014} and the presence of sets of extended states in Hall measurements~\cite{Li:2007ju,Jung:2011dk}. In the presence of single vacancies (SV) and double vacancies (DV), critical states are found to preclude the formation of the usual graphene LLs. Rather, two sets of extended states form at energies different from $E_n$, within each LL. Consequently, at low energy, by tuning the magnetic field and the impurity concentration, a zero energy Hall plateau can be engineered. We extend the present knowledge by characterizing these states following their real-space behavior. We find that the critical states are predominantly located in the impurity dense regions, while the localized states are trapped inside the pristine-like regions. This suggests that the disorder is triggering a percolation of states mechanism, which is also supported by our estimation of the critical exponent. Furthermore, by calculating the transverse conductivity numerically, we prove that $\sigma_{\text{xy}}(E)$ retains quantized values between Landau levels, even in a highly disordered environment. Our numerical approach circumvents the problems commonly associated with Chern number calculation from (i) a computational point of view, namely we don't have to diagonalize matrices containing millions of elements, and (ii) a physical point of view, as the Chern number approach might become ill-defined when the gaps between extended states close due to increasing disorder contributions.

\section{Model and Methods} 

A single-orbital first-neighbor tight-binding (TB) model restricted to $p_z$ orbitals is used to describe graphene, with hopping terms equal to $\gamma_0$ and zero on-site energies. Model SV and DV are described by removing the corresponding carbon orbitals, and are randomly distributed on graphene samples containing up to 12 million atoms. SV and DV are short-range scatterers that entail inter-valley scattering~\cite{PhysRevB.75.033412,PhysRevB.76.205408,OST_PRB77}, but with genuine differences. Indeed, SV locally break the sublattice symmetry and induce stronger localization effects, whereas DV locally preserve the sublattice symmetry. Both of these defect models retain electron-hole symmetry, which simplifies calculations and analysis of the physics at play, unhindered by the full complexity of DFT-fitted models, such as for oxygenated graphene~\cite{2053-1583-1-2-021001}.

The order-N method to obtain the dissipative bulk conductivity by wavepacket evolution is already well established~\cite{Roche19992284,Fan2014}. By following the time evolution of the wavepackets, length-dependent conductivities $\sigma_{\text{xx}}(L)$ can be extracted, probing diffusive and localization regimes. The effect of a perpendicular magnetic field is modeled through a Peierls phase substitution~\cite{Luttinger}. As the spin polarization is neglected in present simulations (a factor two is included to take into account the spin degree of freedom), Zeeman splitting is not considered. We don't expect this to alter our conclusions, as is discussed at the end of the paper. 

The non dissipative Hall conductivity is calculated using a newly developed efficient real-space algorithm~\cite{OrtmannArxiv, Ortmann_2015} [see also Ref.~\cite{PhysRevLett.114.116602}] as follows:
\begin{multline*}
  \sigma_{\text{xy}}(E,t) = - \frac{2}{V} \int_0^\infty dt e^{- \eta t/\hbar } \int_{-\infty}^{\infty} d E^\prime f(E^\prime - E) \\ \Re e \left[ \left< \phi_{\text{RP}} \middle| \delta(E^\prime - \hat{H}) \hat{j}_y \frac{1}{E^\prime - \hat{H} + i \eta} \hat{j}_x (t) \middle| \phi_{\text{RP}} \right> \right]
  \label{}
\end{multline*}
with $V$ the volume of the system, the current operator $\hat{j}_x$, $f$ the Fermi function and $\eta \rightarrow 0$ a small parameter required for numerical convergence. $\left|\phi_{\text{RP}} \right>$ is a random phase state that allows to drastically limit the computation time.

To simulate two-terminal transport, we consider a standard configuration composed of a graphene ribbon with a central region of length $L$, where DV are randomly distributed. To mimic source and drain contacts, graphene is highly doped outside this region. An on-site energy shift of $\gamma_0$ in the Hamiltonian describing the contacts accounts for the doping. We obtain the differential conductance of the system and the spatial distribution of the spectral current by means of the Green's function approach~\cite{Cresti2006}.

\section{Results}
\subsection{Density of States} 

\begin{figure}[t]
\begin{center}
\includegraphics[width=1\columnwidth]{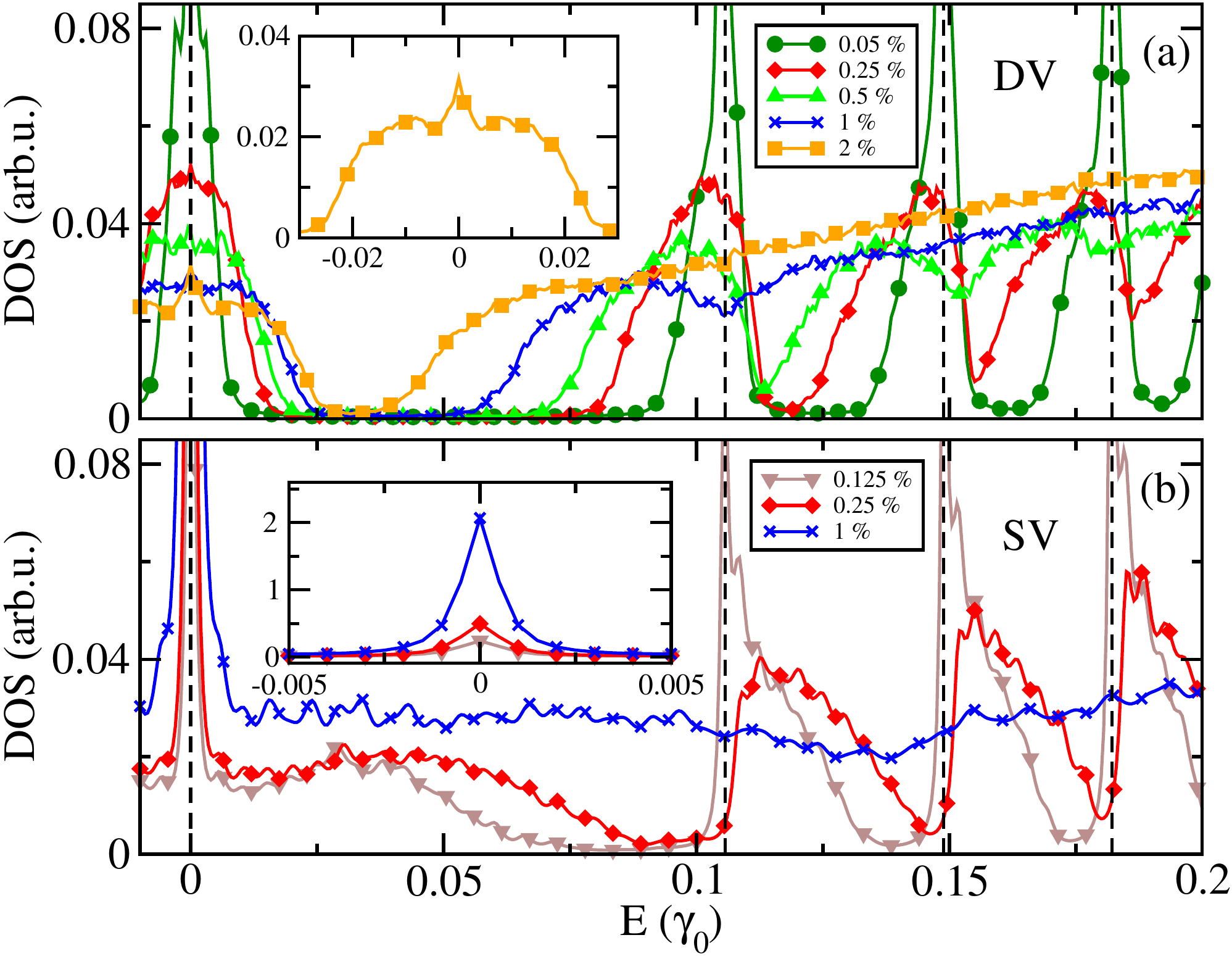}
\caption{\label{fig1}
(color online) DOS for DV (a) and SV (b) at $80$ T, for different impurity concentrations. The curves are symmetric around $E = 0$, so only the electron side is plotted. The dashed vertical black lines indicate the energy positions $E_n$ of LLs in the conventional pristine quantization. Insets provide zooms at low energies.%
}
\end{center}
\end{figure}

As a first step, we calculate the density of states (DOS) for different vacancy densities in the presence of a magnetic field of $80$ T, see Fig.~\ref{fig1}. For the chosen densities, the average distance $d$ between defects is in the order of the magnetic length ($l_B \approx 25$ nm $/ \sqrt{B/T}$), thus favoring strong coupling between impurity states. For the weakest impurity concentrations considered in Fig.~\ref{fig1} ($0.05\%$ for DV and $0.125\%$ for SV), LLs still exist at conventional quantization energies $E_n$. Yet, they are broadened due to lingering coupling between impurities (the transition to the non-coupled case is discussed in Sect.~\ref{levelCondensation}). The LLs at conventional energies gradually disappear for larger density, when impurity states emerge at energies below and above each pristine LL energy $E_n$ for the DV case, and at energies larger than $E_n$ for the SV case, in agreement with the predictions for a (quasi-) periodic model~\cite{PER_PRB78}. In contrast, the robustness of the zero energy states in the SV case in Fig.~\ref{fig1} is explained by the rank-nullity theorem~\cite{Pereira2008}. This theorem predicts the existence of zero-energy modes (ZEM) for dilute concentrations of SV, while they should not form for DV~\cite{Trambly2011}. 

\subsection{Longitudinal and transverse conductivity} 

\begin{figure}[t]
\begin{center}
\includegraphics[width=1\columnwidth]{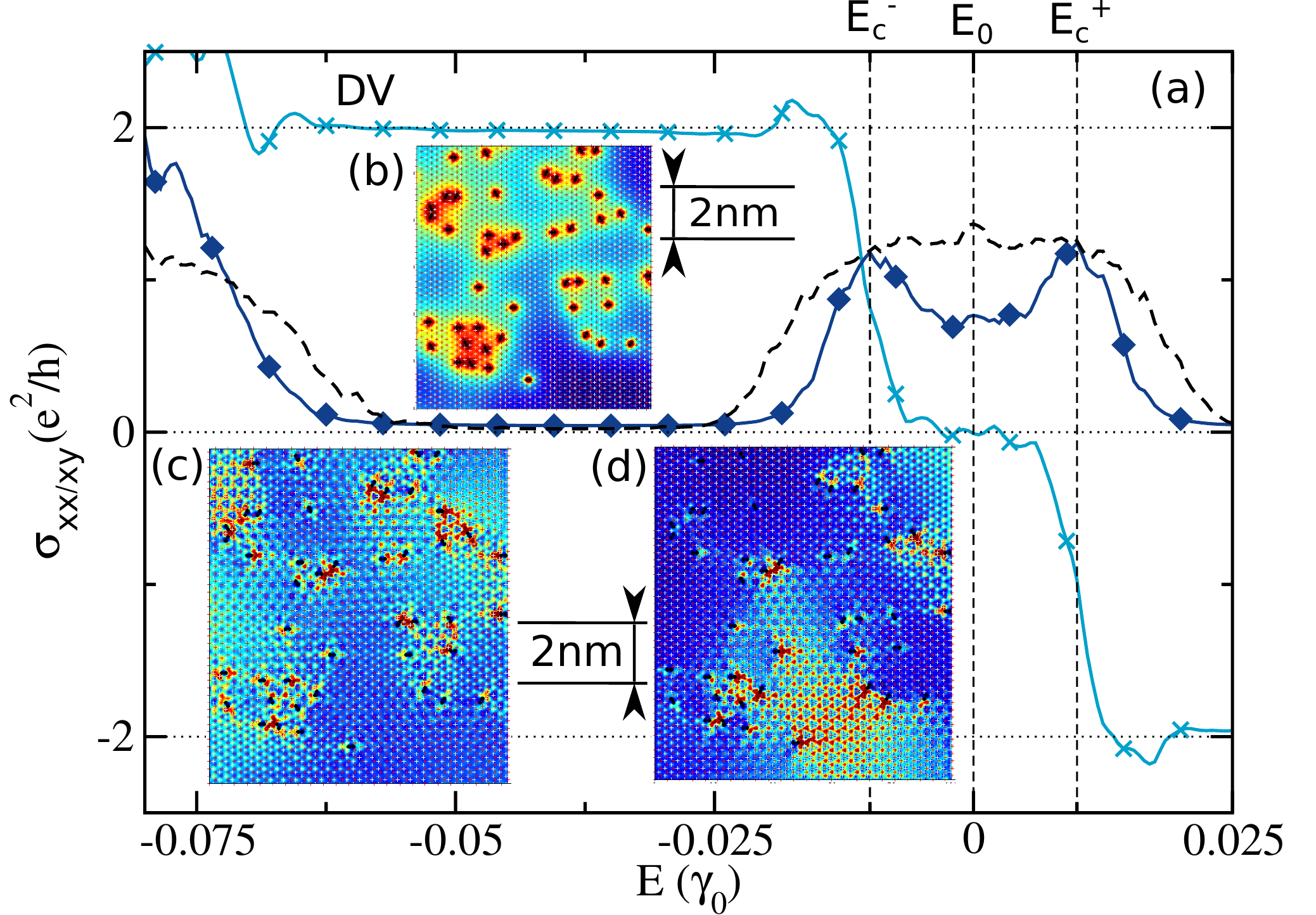}
\caption{\label{fig2}
(color online) $\sigma_{\text{xx}} (E)$ (dark-colored diamonds) and $\sigma_{\text{xy}} (E)$ (light-colored crosses) in compared with the DOS (dashed black lines) for $1\%$ of DV in (a) at $80$T. Horizontal dotted lines give the expected Hall plateaus for IQHE filling factors. Vertical dashed lines in (a) locate the energies $E_c^+$ and $E_c^-$. Inset (c) shows the PDOS for critical states at $E_c^+$ and $E_c^-$, while (d) shows the PDOS for $E_0$. Inset (b) gives the impurity density $W_i$ (see text). 
}
\end{center}
\end{figure}

\begin{figure}[t]
\begin{center}
\includegraphics[width=1\columnwidth]{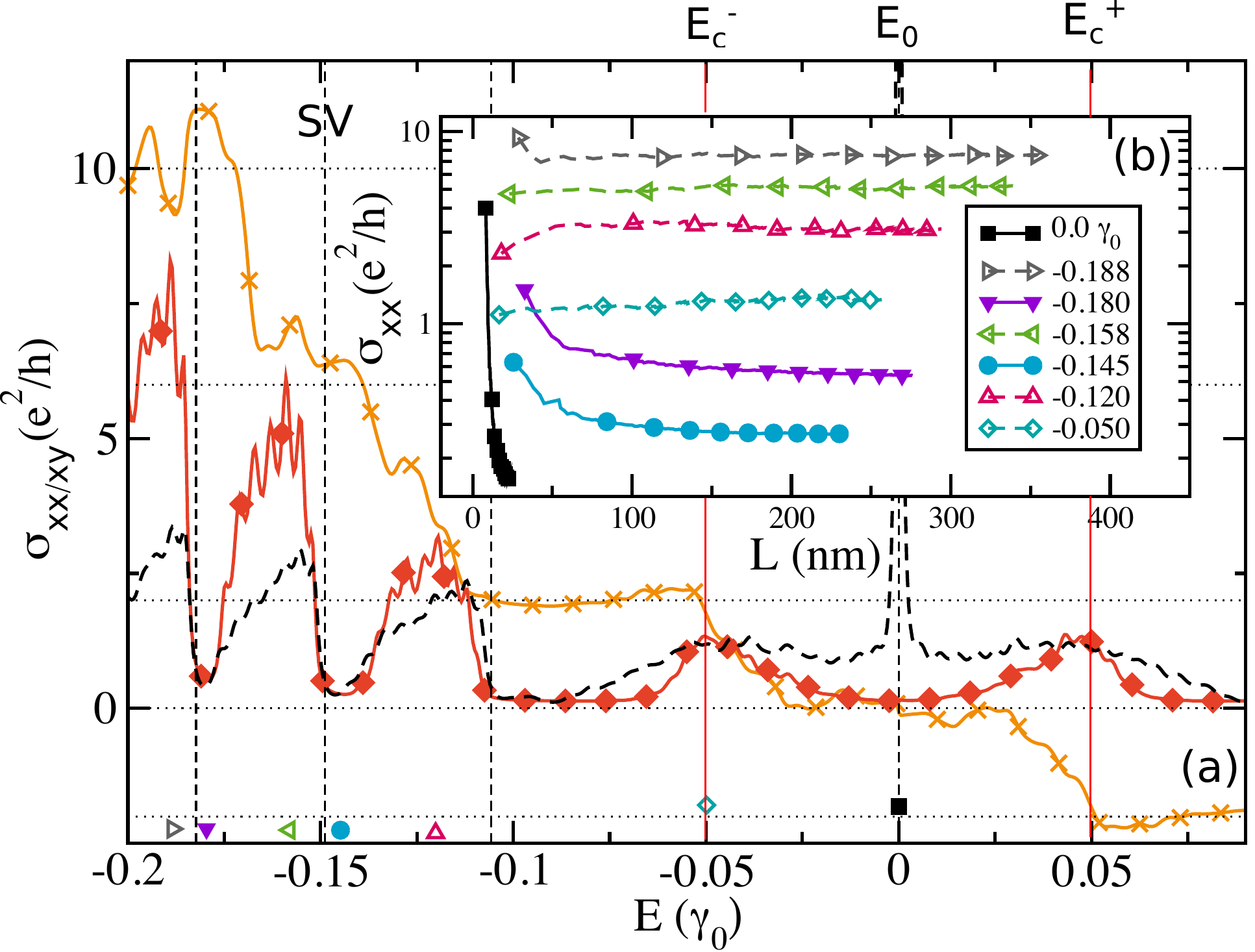}
\caption{\label{fig3}
(color online) $\sigma_{\text{xx}} (E)$ (dark-colored diamonds) and $\sigma_{\text{xy}} (E)$ (light-colored crosses) in comparison with the DOS (dashed black lines) for $0.25\%$ of SV in (a) at $80$T. Horizontal dotted lines give the expected Hall plateaus for IQHE filling factors. Vertical red dashed lines in (a) locate the energies $E_c^+$ and $E_c^-$. The length-dependent conductivities $\sigma_{\text{xx}}(L)$ at selected energies are shown in (b). Solid lines with full symbols correspond to localized energies, while dashed lines with open symbols correspond to extended states. Vertical dashed lines in (a) correspond to conventional $E_n$ energies of pristine graphene. Symbols are only plotted every other tens of data points, for clarity.%
}
\end{center}
\end{figure}

To study the nature of these impurity states, we perform conductivity calculations for $1\%$ of DV, see Fig.~\ref{fig2} (a), and for $0.25\%$ of SV in Fig.~\ref{fig3}(a). By computing both the $\sigma_{\text{xx}} (E)$ (dark color) and $\sigma_{\text{xy}}(E)$ (light color), the occurrence of localized and extended states for different energies is clarified.

Focusing on the low energy region, Figs.~\ref{fig2} and ~\ref{fig3} show that, for both SV and DV, the impurity states of Fig.~\ref{fig1} are more extended at the energies $E_c^+$ and $E_c^-$, where $\sigma_{\text{xx}} (E)$ peaks appear. For DV, the DOS does not exhibit any double peak structure (dashed line in Fig.~\ref{fig2}), while the conductivity clearly resolves it. The different position of the peaks for SV and DV is reminiscent of their behavior when arranged in periodic arrays~\cite{PER_PRB78}.

For both SV and DV, the height of the two peaks ($\sigma_{\text{xx}} (E) \simeq 1.2 e^2/h$) resembles the value for the case where inter-valley mixing leads to two sets of extended states within the same Landau level~\cite{OST_PRB77}. This explains the modified transition between quantized Hall plateaus at $-2 e^2/h$ and $2 e^2/h$, with an additional plateau at $E=0$, observed in Fig.~\ref{fig2} and Fig.~\ref{fig3} for DV and SV, respectively. The difference in strength of DV compared to SV leads to $\sigma_{\text{xy}} (E)$ profiles with varying slope between $-2 e^2/h$ and $2 e^2/h$. Such gradient in the localization strength has also been observed in Ref.~\cite{PhysRevLett.101.036805}, where the considered magnetic fields are several orders of magnitude larger. For the selected DV case in Fig.~\ref{fig2}(a), the slope at $E_0$ is not completely equal to zero, thus still providing a limited contribution to the longitudinal conductivity (around $0.7 e^2/h$ for the calculated length-scale).

The impurity states at $E_c^-$ and $E_c^+$ below and above $E_0$ (position of the conventional zero energy LL) only contribute $2e^2/h$ each to the transverse conductance, indicating that they originate from the original LL0, which would contribute $4 e^2/h$ (spin degeneracy included). Because of the neglected spin polarization in our simulations, these results support the valley-mixing scenario predicted theoretically~\cite{PhysRevB.75.033412,PhysRevB.76.205408,OST_PRB77}, where the existence of the two extended states at $E_c^-$ and $E_c^+$ results from the mixing between $K$ and $K^\prime$ valleys, within the same LL.

To illustrate the spatial distribution of extended impurity states at low energy, we calculate the projected DOS (PDOS) at $E_c^{\pm}$ and $E_0$ in insets (c) and (d) respectively of Fig. 2, and compare it to the local impurity density defined as $W_i = \sum_{j}^{N_{\text{imp}}} 1/d_{ij}$, where $d_{ij}$ is the distance between atom $i$ and impurity $j$, as displayed in inset (b). Both states at $E_c^+$ and $E_c^-$ give exactly the same density plots, emphasizing that they have indistinguishable real space distributions.  Blue regions in (b) are less dense in impurities than the red ones. A clear correlation is observed between the location of the extended states at $E_c^{\pm}$ (c) and the impurity density (b), i.e. the extended states mainly spread over the impurity regions of the sample, while the localized states (d) are bound to impurity-free areas.
The delocalized nature of states at $E_c^{\pm}$ (c) is further confirmed by the strongly reduced maximum in the PDOS ($0.06$ arb.~units) compared to the maximum PDOS for $E_0$ (d) which reaches $0.25$ arb.~units.

In addition, to extend these results to the higher energies and to LLs different from LL0, length-dependent conductivities $\sigma_{\text{xx}}(L)$ are considered, whose decay is related to the strength of localization effects, see Fig.~\ref{fig3}(b). The SV case is chosen as it depicts better energy resolution between extended and delocalized states at high energy. New sets of extended impurity states clearly develop also away from the Dirac point up to $-0.2 \gamma_0$, witnessed by the longitudinal conductivity. Extended state energies in $\sigma_{\text{xx}}(L)$ are confirmed by the plateau transitions in $\sigma_{\text{xy}} (E)$ [Fig.~\ref{fig3}(a)]. Contrary to the LL0 case, extended impurity states at higher energy contribute to $\sigma_{\text{xy}} (E)$ with integer multiples of $4 e^2/h$; no clear step is observed at $\sigma_{\text{xy}} = \pm 4e^2/h$ for instance. Traces of quantization are found at $\pm 6e^2/h$  and $\pm 10 e^2/h$ for SV in Fig.~\ref{fig3}(a) . Similarly to the low energy case, these states form at energies different than the ones predicted from conventional Hall quantization. The full $4 e^2/h$ steps at higher energies (in contrast with the two $2 e^2/h$ steps in LL0) are rationalized by the fact that, even for the fully periodic case where random disorder broadening is absent, the two new states within each higher-energy-LL are very close in energy~\cite{PER_PRB78}. Thus, unrealistically small broadening and a complete absence of disorder would be required to resolve the energy lifting of extended states for LLs different than $LL0$.

Finally, from this $\sigma_{\text{xx}}(L)$ plot, we also note that, although SV significantly contribute to the DOS at $E_0$, these states (ZEM) turn out to be strongly localized, in sharp contrast with the DV case where the conductivity at $E=0$ remains finite for much larger length scale. In the next section, this different behavior is further scrutinized by simulating different densities.

\subsection{Coupled impurity states} 

\begin{figure}[t]
\begin{center}
\includegraphics[width=1\columnwidth]{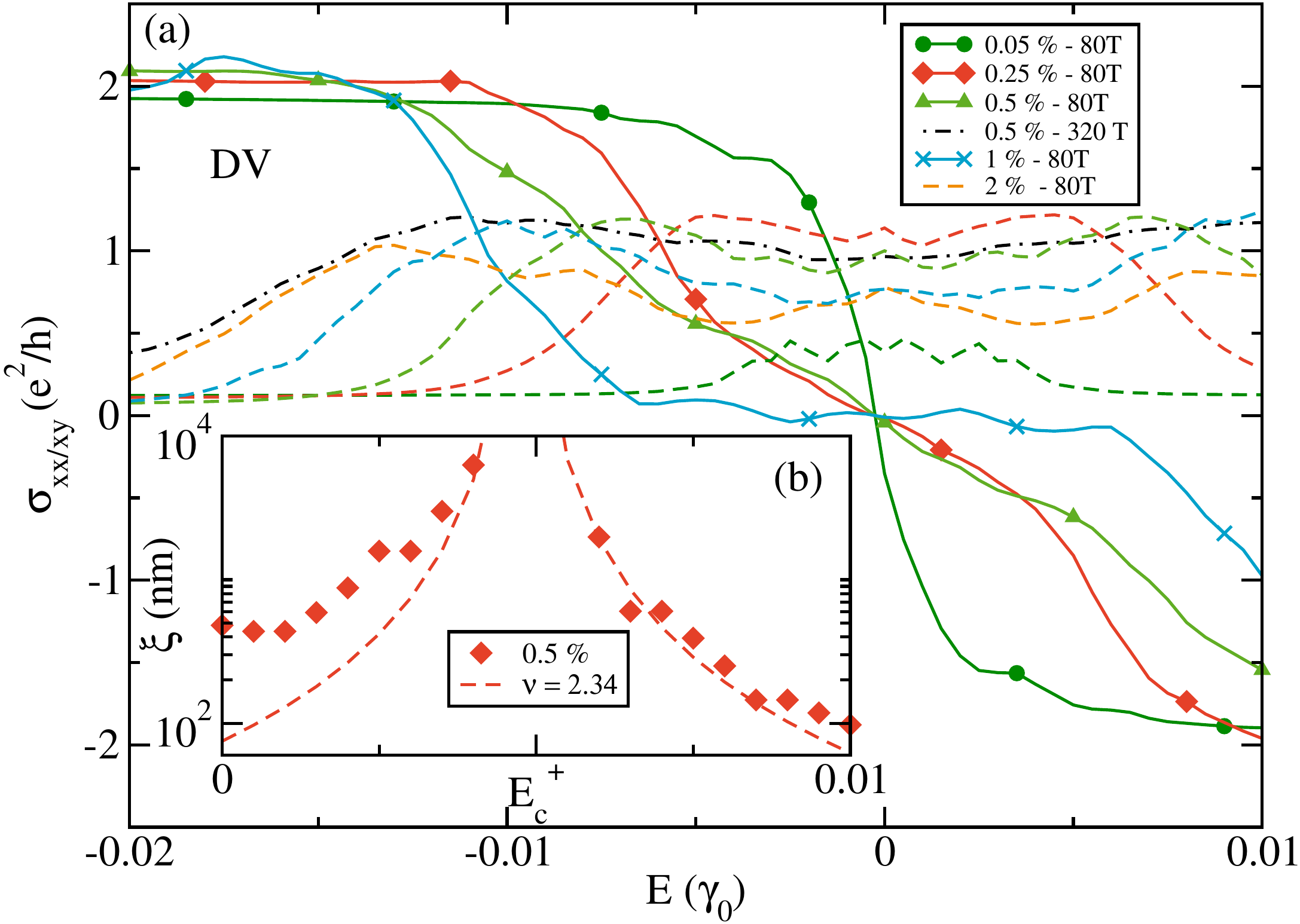}
\caption{\label{fig4}
(color online) $\sigma_{\text{xx}} (E)$ (dashed lines) and $\sigma_{\text{xy}} (E)$ (solid lines) for DV at $80$ T (a). $\sigma_{\text{xx}} (E)$ is also calculated for $0.5\%$ at $320$ T and $2\%$ at $80$ T.  Estimated localization lengths for $0.5\%$ at $80$ T with theoretical critical exponential ($\nu=2.34$) decay around $E_c^+$ (b). 
}
\end{center}
\end{figure}

\begin{figure}[t]
\begin{center}
\includegraphics[width=1\columnwidth]{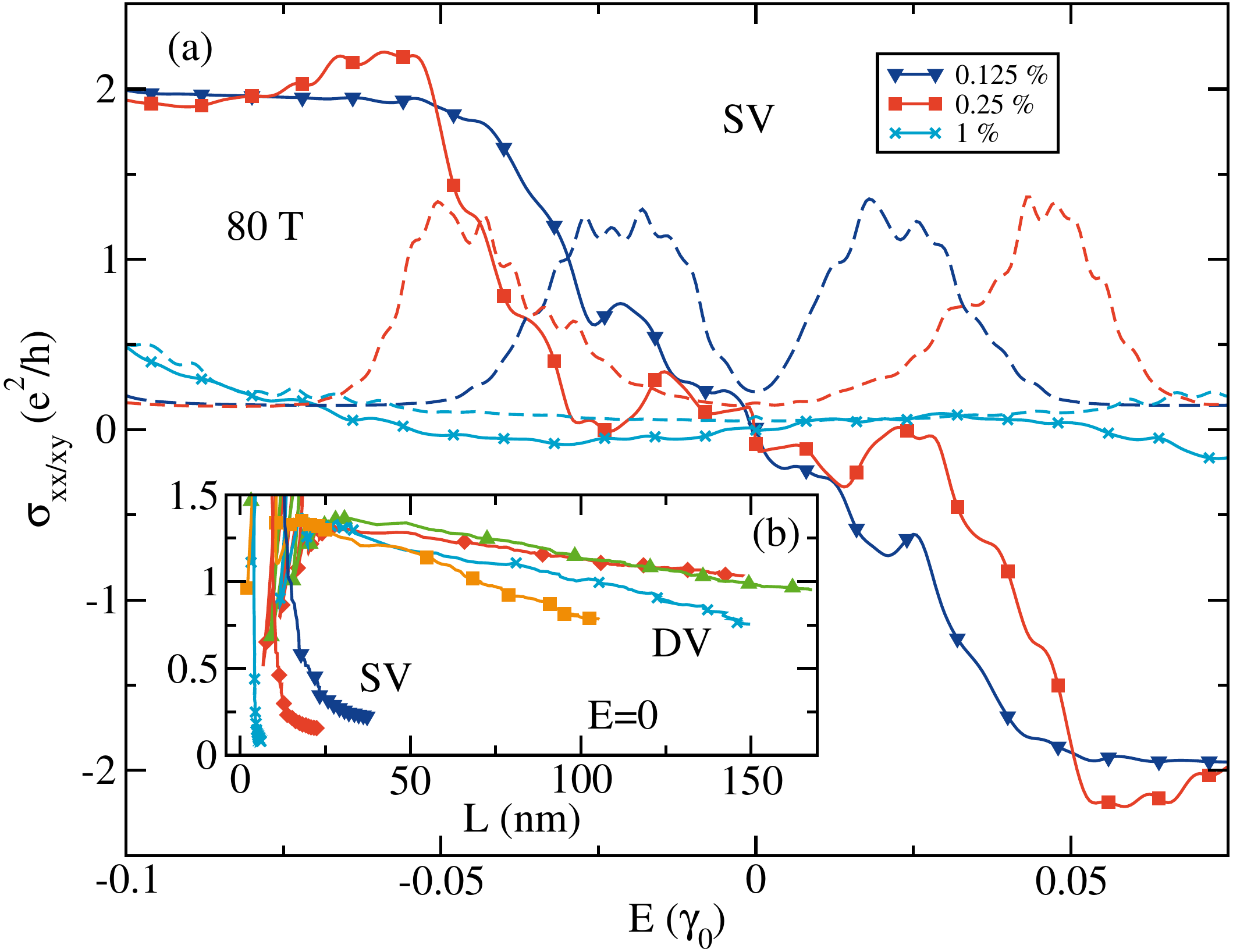}
\caption{\label{fig5}
(color online) (a) $\sigma_{\text{xx}}(E)$ (dashed lines) and $\sigma_{\text{xy}}(E)$ (solid lines) for  SV at $80$ T (a). (b) $\sigma_{\text{xx}}(L)$ for $E_0$ for both SV and DV. Symbols are only plotted every other tens of data points, for clarity. $y-$axis label (b) is same as (a).
}
\end{center}
\end{figure}

In the very dilute limit (around $0.05\%$), localized impurity states can be sufficiently separated to have no significantly overlap. In this limit, the conventional QHE is essentially preserved~\cite{PhysRevLett.113.186803}. This limiting case will be considered separately in Sect.~\ref{levelCondensation}. By increasing the impurity density, noticeably extended states start to appear when $l_B$ is of the order of $d$ as in the case for $1\%$ of DV in Fig.~\ref{fig2}(a) and for $0.25\%$ of SV in Fig.~\ref{fig3}(a), when localized impurity states couple. This condition can be satisfied by increasing the impurity concentration or the magnetic length (by decreasing the magnetic field), which we demonstrate for several cases in Fig.~\ref{fig4}(a) and Fig.~\ref{fig5}(a). Once in the coupled regime, the energy of the extended states increases with $B$~\cite{PER_PRB78,BAH_PRB79} and concentration, see for instance Fig.~\ref{fig4}(a) for 0.5\% DV concentration at $320$ T.

To analyze the percolation of states driven by impurities, the extracted localization lengths in the strong localization regime, using~\cite{Leconte2011} 
\begin{equation}
\sigma_{\text{xx}}(L) \sim \exp{\left[-\frac{L}{\xi} \right]},
\end{equation}
are plotted in Fig.~\ref{fig4}(b) for an impurity concentration of $0.5\%$ (symbols) (similar results are obtained for other concentrations, not shown here). Percolation theory~\cite{0022-3719-21-14-008,PhysRevB.75.033412} predicts a critical exponent $\nu=2.34$ for $\xi \sim |E-E_c|^{-\nu}$ (solid line). Visual agreement is obtained between numerics and theory for the right tail of $E_c^+$. However, for the left tail, towards $E_0$, agreement cannot be claimed. This behavior results from the set of remnant states in the low impurity density regions of the samples, which are not fully localized for DV at the considered length scale. To further characterize the puzzling behavior of states at $E=0$, we plot $\sigma_{\text{xx}}(L)$ in Fig.~\ref{fig5}(b) for SV and DV. On one hand, for SV the states are strongly localized. This is in agreement with the localization behavior of ZEM predicted at zero magnetic field~\cite{PhysRevLett.110.196601,PhysRevLett.113.186802}, following a power-law behavior $\sigma_{\text{xx}}(L) \sim L^{-2}$. On the other hand for the DV case, $\sigma_{\text{xx}}(L)$ exhibits a linear decay. This explains the finite conductivity contributions observed in Fig.~\ref{fig2}(a) and Fig.~\ref{fig4}(a). Actually, the highest density concentration ($2\%$), with an increased energy split between extended impurity states, even allows resolving the three sets of states at $E_c^{\pm}$ and $E_0$ in the $\sigma_{\text{xx}} (E)$ curve. This is counterintuitive in the sense that increasing the disorder decreases the amount of clean patches in the sample, the habitat for delocalized states at $E_0$. $\sigma_{\text{xx}} (E)$ at $E_0$ remains nevertheless surprisingly robust up to long length scales, which is not observed for SV. This could either be explained by the weaker DV disorder strength allowing states to propagate more easily from one pristine patch to the other, or by referring to Ostrovski~\textit{et al.}~\cite{PhysRevLett.113.186803,PhysRevB.85.195130} providing an argument against localization at $E=0$ for point-like chiral disorder. Ref.~\cite{Ferreira_2015} recently demonstrated through simulations that it is very difficult and computationally much more demanding to accurately capture the singularity associated to the ZEM, even more so for stronger SV compared to DV. We thus remain cautious about making conclusions on the exact localization behavior at the $E=0$ point.


\subsection{Two-terminal Calculations}

\begin{figure}[t]
\begin{center}
\includegraphics[width=1\columnwidth]{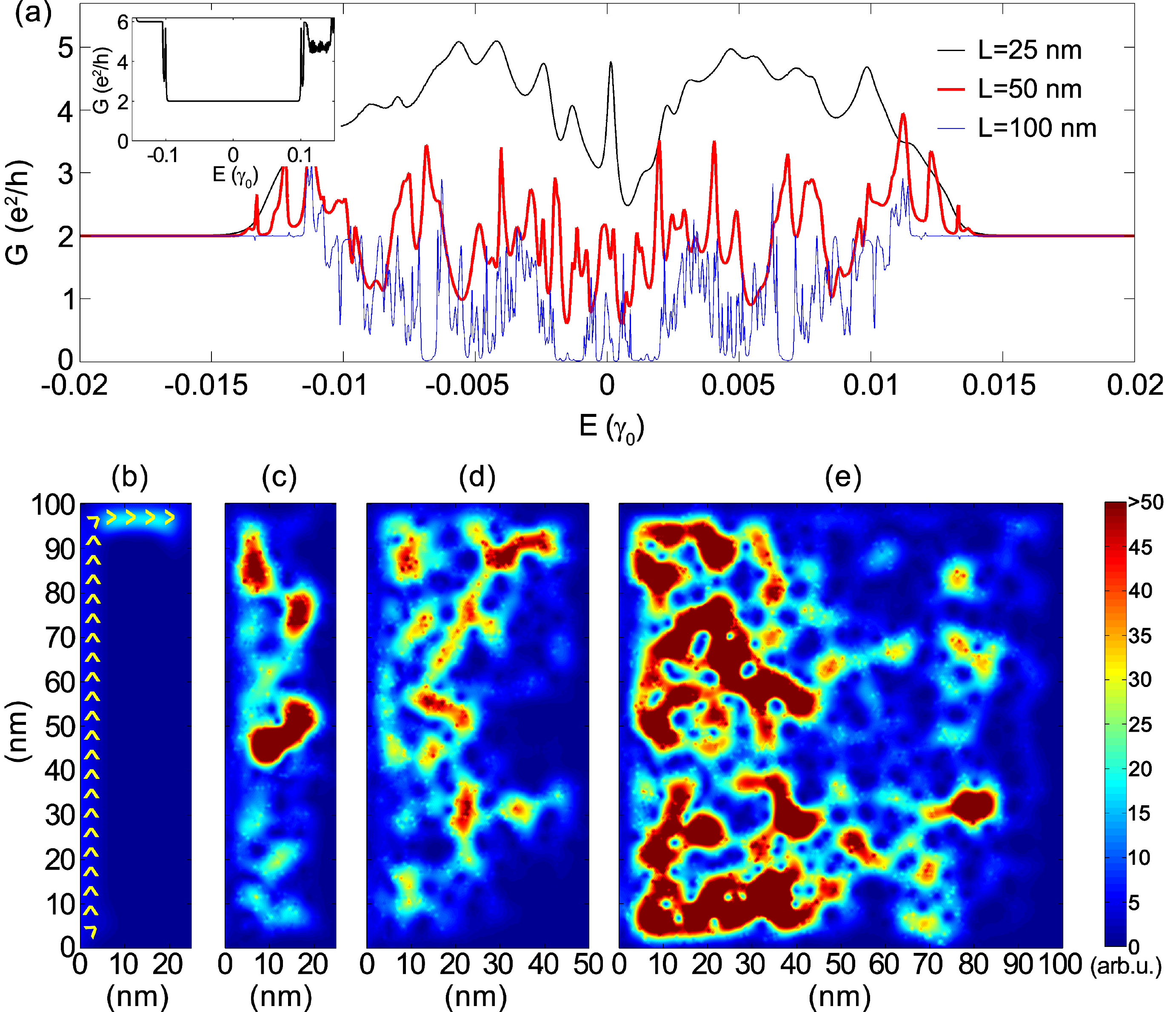}
\caption{\label{fig6}
(color online) (a) Conductance of the two-terminal system with a density $n$=0.5\% of DV over a ribbon of length $L$ for 25 nm to 100 nm. Inset: Pristine case for $L$=25 nm. (b) Spatial distribution of the spectral current at $E$=0.008 $\gamma_0$ for the pristine ribbon with $L$=25 nm. The arrows indicate the current direction. (c-e) Same as (b) for DV density of 0.5\% and $L$=25 nm, 50 nm and 100 nm. All simulations at $80$ T.%
}
\end{center}
\end{figure}

To gain complementary insight in the length scaling and percolation behavior of impurity states, we also use a different methodology,  namely by calculating the two-terminal conductance of a $100$ nm wide ribbon for $B=80$ T. In the pristine case, the conductance shows a $2e^2/h$ plateau in the energy region of Fig.~\ref{fig6}(a) (see the inset for a larger energy range). The corresponding spectral current injected from the left contact and transmitted along the chiral top edge channel of the ribbon is indicated by the arrows in Fig.~\ref{fig6}(b). For Figs~\ref{fig6}(c)-(e), $0.5\%$  of DV  are distributed over a ribbon of length $L$ from $25$ nm to $100$ nm. In quantitative agreement with above calculations in 2D geometry, Fig.~\ref{fig6}(a) shows that DV induce bulk states within a certain energy window (from $-0.015 \gamma_0$ to $0.015 \gamma_0$), which modify the pristine conductance. 
For $L$=25 nm, most of these states are extended enough to connect source and drain contacts and induce a conductance increase well above the pristine value. 
In fact, the observed conductance for this case is reminiscent of the DOS curve in Fig.~\ref{fig1}(a) and the bulk spectral current distribution in Fig.~\ref{fig6}(c) illustrates in real space that electrodes are connected by the states and explains the high conductance also seen in 2D simulations.
When increasing $L$ to $50$ nm, less bulk states are sufficiently extended to allow the electrons reaching the drain contact. As a consequence, the conductance decreases and narrow peaks appear corresponding to the energy regions where more extended states are concentrated, contributing to the transport. The current distribution of Fig.~\ref{fig6}(d) indicates that, for this energy, the electron penetration decreases, with a less efficient bridging between electrodes. Analogous behavior is observed for $L=100$ nm in Fig.~\ref{fig6}(e), with a more pronounced fragmentation into peaks and a weaker conductance decrease at the center and the sides of the DV energy window [Fig.~\ref{fig6}(a)], in agreement with the 2D simulations. For $L>100$ nm, the peaks reduce to isolated resonances with conductance below or almost $2e^2/h$, and a transport gap progressively opens around $E=0$ (not shown here).

Note that, in the 2D results, the narrow resonances are masked by self-averaging effects. On the other hand, two-terminal simulations should be performed over a large ensemble of disorder realizations to recover the statistical information provided by 2D bulk conductivity simulations, such as the exact position of the more extended states corresponding to the critical energies.
The broadness of energy distribution of the DV-induced states and their localization is inversely proportional to the defect density. In the limit of a periodic DV distribution~\cite{PER_PRB78}, the bulk states are completely delocalized and concentrated around very specific energies.

\subsection{Level-condensation}
\label{levelCondensation}

In the previous sections, we have always considered a distance between defects short enough to allow their coupling and the formation of impurity states at new critical energies. In Ref.~\cite{PhysRevLett.113.186803}, the transition from interacting impurities to non-interacting impurities in the very low concentration regime has been considered. A distance criterion is provided at which so-called level-condensation should occur, namely
\begin{equation}
  r = \frac{l_B}{l_\text{imp}} < 0.39.
  \label{}
\end{equation}
For the case of $80$T and the lowest concentration considered so far ($0.05 \%$), $r = 0.63$. By reducing the concentration to $0.01\%$, one gets $r=0.28$, for which level condensation should occur. Similarly, by increasing the magnetic field, and keeping $0.05\%$ of DV, one can achieve values of $r=0.32$ (for $320$T) and $r=0.16$ (for $1280$T), respectively. The latter approach (high magnetic field - low concentration) is simpler from a computational point of view, the former (lower magnetic field - higher concentration) is more realistic from an experimental point of view.

\begin{figure}[t]
\begin{center}
\includegraphics[width=1\columnwidth]{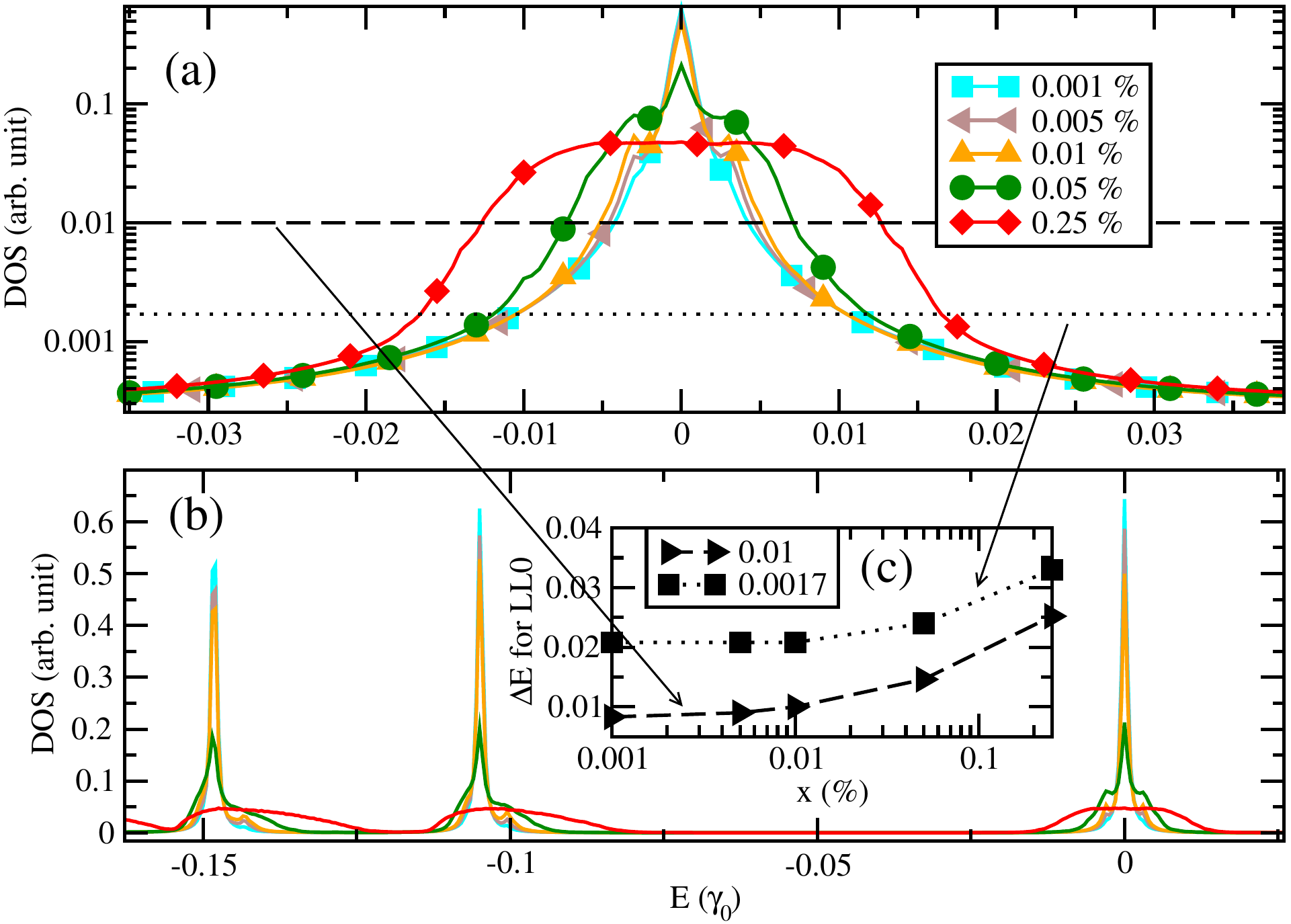}
\caption{\label{fig7}
(color online) DOS around the LL0 for DV (a). DOS for a larger energy spectrum (b). (c) gives the energy width of LL0 at selected heights [see dotted and dashed line in (a)], for different impurity concentrations.%
}
\end{center}
\end{figure}

When calculating the DOS, very small impurity concentrations can be considered, as reported in Fig.~\ref{fig7}. Panel (b) depicts a large energy range, while panel (a) focuses on the energy region around the $LL0$. A logarithmic scale was used for the zoom, for the sake of clarity. With the broadening energy set to $0.0005 \gamma_0$, the concentration-dependent width of the peak ($\Delta E$) is reported in panel (c) for two arbitrary peak heights. For the dotted line ($0.0017$ arb. unit), the value of the splitting does not vary for concentrations below $0.01 \%$. This confirms qualitatively and quantitatively (with an uncertainty related to height and numerical broadening) the level condensation from a DOS point of view. A conductivity analysis for the \textit{condensated} regime is out of the scope of the present study, because all states become localized due to the magnetic field. Either edges or additional long-range disorder would have to be included to allow for extended states to develop at the energies $E_n$ of LLs in the conventional pristine quantization.

\section{Conclusion} 

We have investigated numerically the possible origin of anomalous features reported in the quantum Hall regime of low mobility graphene samples~\cite{Nam:2013ey,Nam2014,Li:2007ju}, such as resonances in the dissipative conductivity and a zero-energy Hall plateau. They result from the formation of disorder-induced percolating bulk states, whose density and extension is maximal around two critical energies that depend on the magnetic field and on the impurity density. The presence of defect-induced critical states on novel QHE properties does not seem to depend on the local symmetry breaking (as induced by SV), as they also form for DV impurities. Rather, valley-mixing is required. Finally, a vanishing contribution of these defect-induced critical states is observed in the very low impurity density limit, characterized by so-called \textit{level condensation}~\cite{PhysRevLett.112.026802}. 
Results on highly disordered graphene using more realistic impurities, with TB parameters extracted from \textit{ab initio} simulations, suggest a certain universality of these impurity-induced extended states~\cite{2053-1583-1-2-021001, PhysRevB.78.165403,wei2012nature}, even in the presence of electron-hole asymmetry. This asymmetry can provide additional signature on the nature of the contaminating impurities~\cite{Bai_2015}. The energy
dependence of impurity states with magnetic field can be inferred from existing DOS literature for different types of impurities~\cite{Yuan_2010, PER_PRB78, 2053-1583-1-2-021001}. The study of electron transport, as we performed in this paper, is nevertheless required to precisely assess the extendedness of states, as is apparent from the energetically unresolved static DOS features for the DV case. 

The inclusion of a Zeeman term in the present form of our Hamiltonian, while desirable for high magnetic fields, is not expected to alter our conclusions, as we do not consider spin-spin or spin-orbit interactions at this point. Such term would then simply induce two spin-dependent copies of the same physics (and an additional trivial splitting with the Zeeman energy). We also note that the present conclusions at high magnetic fields ($80$ T in this work), which are computationally less demanding for numerical convergence in the transverse conductivity than at low magnetic fields, are expected to be robust for much smaller magnetic fields (where Zeeman interaction is weakened), as is demonstrated in a separate work on oxygenated graphene~\cite{2053-1583-1-2-021001}. In our work, we demonstrate that the unconventional transport features can be explained even with this simplification in neglecting the Zeeman term. 
Finally, we propose two future lines of research. First, the way the Chern number is modified or not in highly disordered graphene should be investigated (as well as simply explore if Chern number classification is still appropriate). Our method has the advantage to predict the Hall conductivity even for large disorder, but we presently do not have any tool to calculate the Chern number without going through exact diagonalization of a system containing millions of elements. Second, we comment on how interaction effects might play a role. Indeed, high magnetic fields may induce strongly localized states. The interaction between particles may thus become more relevant. Possible influences could be an interaction-induced localization of the critical states or an additional splitting as in the case of quantum Hall ferromagnetism~\cite{Young:2012bn}; or a competition between both mechanisms.

\acknowledgments

We acknowledge PRACE for awarding us access to the Curie supercomputing center based in France. This research is partly
funded by the European Union Seventh Framework Programme under Grant Agreement No. 604391 Graphene Flagship. ICN2 acknowledges support from the Severo Ochoa Program (MINECO, Grant SEV-2013-0295). F.O. would like to thank the DFG for financial support (grant OR 349/1-1).

\bibliography{converted_to_latex.bib%
}

\end{document}